\documentclass[lettersize,journal]{IEEEtran}
\usepackage{amsmath,amsfonts}
\usepackage{algorithmic}
\usepackage{algorithm}
\usepackage{array}
\usepackage[caption=false,font=normalsize,labelfont=sf,textfont=sf]{subfig}
\usepackage{textcomp}
\usepackage{stfloats}
\usepackage{url}
\usepackage{verbatim}
\usepackage{graphicx}
\usepackage{cite}
\usepackage{url}
\usepackage{subfig}
\graphicspath{ {figs/} } 
\begin{document}

\title{Decentralized Physical Infrastructure Network (DePIN): Challenges and Opportunities}

\author{Zhibin Lin, Taotao Wang, Long Shi,  Shengli Zhang and Bin Cao
%~\IEEEmembership{Staff,~IEEE,}
        % <-this % stops a space
\thanks{Z. Lin, T. Wang and S. Zhang are with the College of Electronic and Information Engineering, Shenzhen University, China (linaacc9595@gmail.com, ttwang@szu.edu.cn, zsl@szu.edu.cn). L. Shi is with the School of Electronic and Optical Engineering, Nanjing University of Science and Technology, China (slong1007@gmail.com). B. Cao is with the State Key Laboratory of Networking and Switching Technology, Beijing University of Posts and Telecommunications, China (caobin@bupt.edu.cn). \emph{Corresponding Author: Bin Cao}. 
}% <-this % stops a space
%\thanks{Manuscript received April 19, 2021; revised August 16, 2021.}}
}

% The article headers
%\markboth{Journal of \LaTeX\ Class Files,~Vol.~14, No.~8, August~2021}%
%{Shell \MakeLowercase{\textit{et al.}}: A Sample Article Using IEEEtran.cls for IEEE Journals}

%\IEEEpubid{0000--0000/00\$00.00~\copyright~2024 IEEE}
% Remember, if you use this you must call \IEEEpubidadjcol in the second
% column for its text to clear the IEEEpubid mark.

\maketitle

\begin{abstract}
The widespread use of the Internet has posed challenges to existing centralized physical infrastructure networks. Issues such as data privacy risks, service disruptions, and substantial expansion costs have emerged. To address these challenges, an innovative network architecture called  Decentralized Physical Infrastructure Network (DePIN) has emerged. DePIN leverages blockchain technology to decentralize the control and management of physical devices, addressing limitations of traditional infrastructure network. This article provides a comprehensive exploration of DePIN, presenting its five-layer architecture, key design principles. Furthermore, it presents a detailed survey of the extant applications, operating mechanisms, and provides an in-depth analysis of market data pertaining to DePIN. Finally, it discusses a wide range of the open challenges faced by DePIN.
\end{abstract}


\section{Introduction}
\IEEEPARstart{I}{n} the context of the digital age, technological advances and the popularization of the Internet have combined to drive the development of society. However, this process also brings challenges to the existing centralized physical infrastructure networks, which mainly include the risk of data privacy leakage, the possibility of service disruption, and the substantial costs of network expansion. To effectively address these growing issues, an innovative network architecture---Decentralized Physical Infrastructure Network (DePIN)---has emerged. DePIN provides a potential solution to the constraints of conventional centralized infrastructure networks by integrating blockchain technology to decentralize the control and management of physical devices.  

DePIN, a combination of Web 3.0 and Intelligence of Things (IoT), was initially given the name ``MachineFi'' by IoTeX, a innovative blend of the concepts of machine and Decentralized Finance (DeFi) to reflect the potential of machines and their data for financialization. In July 2022, Lattice made an effort to redefine the field by suggesting the term ``TIPIN'' (Token Incentivized Physical Networks). This term was proposed to provide a more precise description of the utilization of token incentives in enabling the joint deployment and operation of a physical network model. In November 2022, Messari launched a poll\footnote{\url{https://twitter.com/MessariCrypto/status/1588938954807869440?ref=iotex.io}} via the Twitter platform to select a term that best represents the field from several candidate names. In this poll, the name ``DePIN'' came out on top with 31.6\% of the votes and was eventually recognized as the official nomenclature of the domain. This marks the formal acknowledgment of DePIN as a definitive term in academic and industrial discourses concerning decentralized physical infrastructure networks.

Compared to the traditional infrastructure networks, DePIN embodies a physical infrastructure network with the following distinguishing features.
\begin{itemize}
\item \textbf{Collective Ownership}: DePIN encourages network participants to actively deploy physical devices to infrastructure networks through token incentives mechanism, establishing a bottom-up self-organizing model. This approach ensures widespread distribution of network ownership, preventing it from being monopolized by a few shareholders or centralized entities.
	
\item \textbf{Low operating costs}: The self-organized model aligns with market economy principles, where resource management within the network is based on incentive mechanisms that maintain a balance between supply and demand, effectively preventing resource waste \cite{li2022blockchain}. Additionally, network contributors bear the responsibility of maintaining their equipment, significantly reducing the overall maintenance costs. Furthermore, the transparent and democratic governance system further reduces unnecessary operating expenses caused by corruption in centralized systems.

\item \textbf{Privacy and Security}: The decentralized architecture of DePIN effectively eliminates the risks of service disruptions, enhancing its overall security. 
In the event of a node or component failure, the remaining network components can sustain to provide service. Furthermore, by leveraging blockchain as the underlying technology \cite{lv2021analysis}, the system safeguards personal data against unauthorized manipulation or access by third parties.
	
\item \textbf{Openness and Innovation}: The inherent decentralization of DePIN substantially reduces market barriers, enabling new entrants to challenge industries that have historically been dominated by a limited number of firms. This heightened openness fosters a competitive environment, spurring innovation and driving the development of superior products and services for consumers. 
\end{itemize}
Therefore, DePIN is considered to have great potential in reshaping many respects of traditional physical infrastructure networks. For example, DePIN's decentralized nature can enhance the trust and security of 6G mobile networks \cite{6g}. By distributing data and computational tasks across multiple nodes, it reduces the risk of single points of failure and improves resilience against cyber threats for 6G networks. Additionally, blockchain technology can be leveraged within DePIN to provide tamper-proof and transparent data records, enhancing the security and trustworthiness of 6G networks.

This article provides a comprehensive exploration of DePIN, presenting its five-layer architecture and key design principles. Moreover, the article provides an in-depth exploration of the existing applications and operational mechanisms of DePIN, and presents market data for DePIN spanning nearly a year. Finally, it concludes with a discussion of the multiple challenges faced by DePIN. The remainder of this article is organized as follows: Section II introduces the DePIN architecture through its five foundational layers. Section III elucidates the key design principles guiding the architecture. Section IV delves into the investigation of the DePIN projects, encompassing its applications, operational mechanisms, and market data analysis. Section VI highlights the challenges facing DePIN. Finally, Section VII concludes the article and suggests directions for future research.

\section{DePIN ARCHITECTURE}
\begin{figure*}
    \centering
    \includegraphics[width=0.8\textwidth]{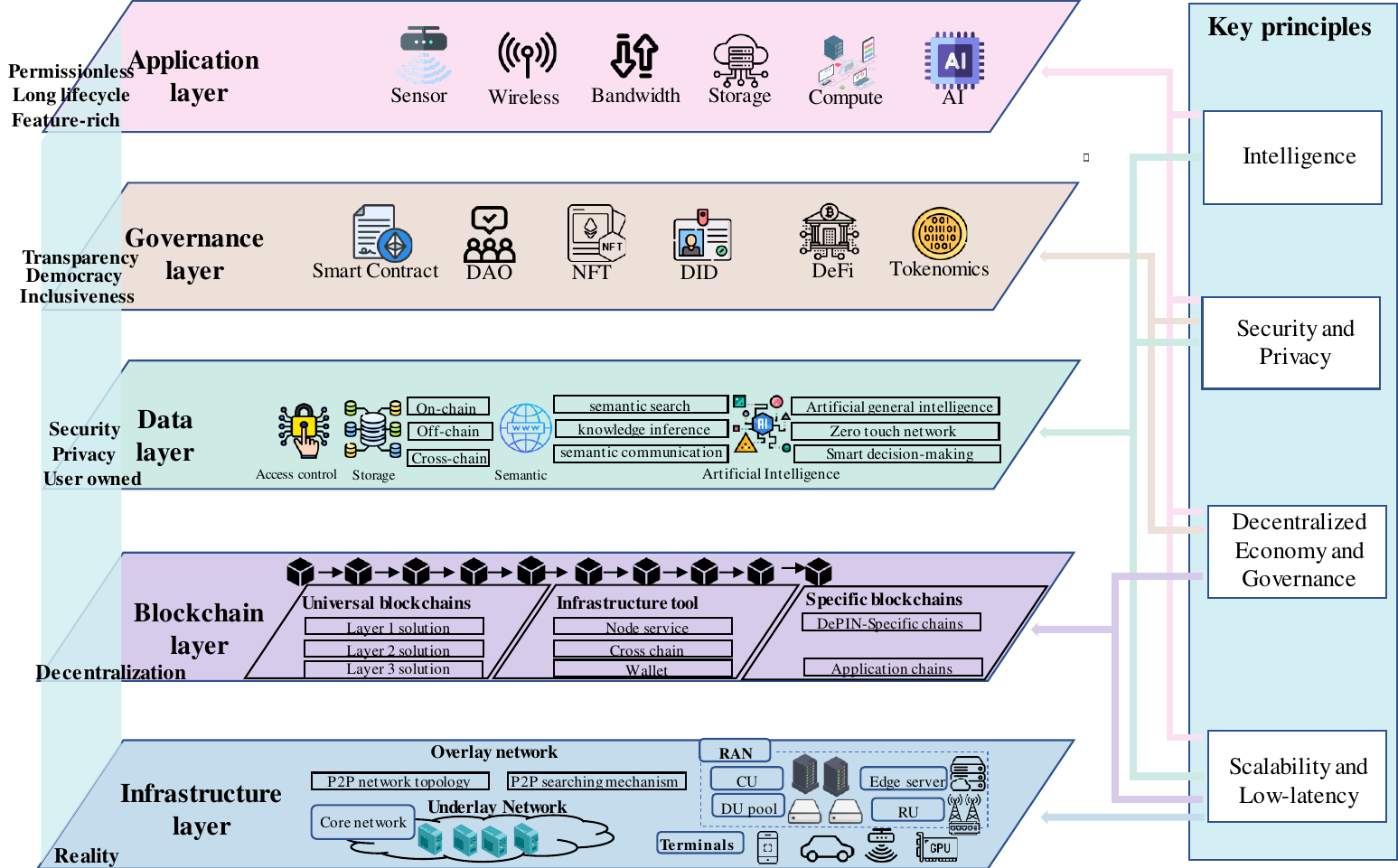}
    \caption{The architecture of DePIN, including the application layer, the governance layer, the data layer, the blockchain layer, and the infrastructure layer}
    \label{architecture}
\end{figure*}

In this section, we provide a comprehensive exposition of the proposed architecture for DePIN. As Illustrated in Fig. \ref{architecture}, the architecture, based on different functions, is intricately structured as a five-layer hierarchy, comprising the application layer, governance layer, data layer, blockchain layer, and infrastructure layer.  

\subsection{Application layer}
DePIN's application layer is meticulously designed to offer a distinct experience compared to Web 2.0 application services, such as license-free access, long-term lifecycle, and feature richness. Facilitated by the decentralized governance layer, application layer achieves permissionlessness, enabling users to freely access the network. Participants can acquire services, and deploy applications without the need for review or approval from any centralized authority. Unlike traditional Web 2.0 applications that rely on the operations of a single company, the application layer's long lifecycle significantly enhances its reliability \cite{sustainability}. In the Web 2.0 model, applications' data and devices are highly centralized, subjecting them to the control of a single company, which can lead to service instability and the risk of applications disappearing. In contrast, DePINs are built upon the governance layer's economic incentives and data layer's data persistence storage capabilities. Thus, DePIN applications will keep serving users as long as there are service providers in the network. Moreover, the application layer exhibits remarkable functional richness, supporting a diverse array of applications and services, including data storage and transmission, resource sharing, and distributed computing. In Section IV, DePIN applications and their categorizations will be explored in further detail.
\subsection{Governance layer}
The governance layer assumes a critical role in establishing the operational rules and institutional foundation of DePIN. It encompasses decision-making mechanisms, transaction rules, incentive mechanisms, and conflict resolution. the governance layer adheres rigorously to the principles of decentralization and privacy protection. By leveraging several core innovations based on smart contracts, such as Non-Fungible Tokens (NFTs),  Decentralized Identifiers (DIDs), Decentralized Autonomous Organizations (DAOs), and DeFi, the governance layer constructs a transparent, democratic, and adaptable governance system \cite{governance}. Supported by the blockchain layer, the governance layer ensures the transparency and auditability of all transactions, enabling participants within the network to observe decision-making processes and resource allocation. Within the governance framework of DAO \cite{dao}, users have equal opportunities to partake in decision-making, effectively preventing the monopolization of decision-making power by a privileged few, embodying the essence of democratic governance. Moreover, the governance layer demonstrates remarkable adaptability, promptly responding to environmental changes and following user needs. This is because of the decentralized decision-making process and financial system, which liberates the governance layer from the constraints imposed by centralized institutions.
\subsection{Data layer}
The data layer fulfills a crucial role in managing the complete lifecycle of all data within DePIN, encompassing data access, storage, and processing. The design of the data layer is guided by a set of core principles, including security, privacy protection, intelligence, and high scalability. First, data access in the data layer involves access authentication and control, protecting the data privacy and ownership of users, through advanced encryption and decentralized identity technologies. Second, the decentralized storage technology, which exploits number of  independent nodes to jointly maintain user data based on a consensus protocol, exhibits greater robustness against cyber attacks compared to traditional centralized storage techniques. It can achieve permanent data storage through a token incentive mechanism. The adoption of decentralized storage technology within the data layer significantly enhances data security and long-term preservation. Regarding data processing, the data layer leverages Semantic Technology \cite{liu2023Web3} to establish a comprehensive system encompassing Semantic Search, Knowledge Inference, and Semantic Communication. This system aims to deliver more precise and personalized search results, deeper data insights, and a more natural and intuitive user interaction experience. Facilitated by semantic technologies, the data layer empowers intelligent data services, such as Artificial General Intelligence, Zero-Contact Networks, and Smart decision-making. These services automate DePIN management and operations while providing accurate and timely decision support. The effective functioning of the data layer also relies on the support of other layers, such as the governance layer providing DID services to ensure secure and reliable user authentication \cite{asn}, the blockchain layer offering trust enhancement and decentralized storage for data, and the infrastructure layer providing physical resource support for data computation and transmission. This cross-layer synergy establishes a robust foundation for managing the complete lifecycle of data within DePIN.  

\subsection{Blockchain layer}The blockchain layer within DePIN’s architecture encompasses blockchain technologies, including general-purpose blockchain solutions, specific blockchain solutions, and blockchain infrastructure tools. The general-purpose blockchain solution encompasses Layer 1 (the base blockchain layer), Layer 2 (the scalability extension solution), and Layer 3 (the application layer), forming a comprehensive blockchain technology stack that ensures seamless service from the base layer to the application layer while addressing scalability challenges. The specific blockchain solutions within the blockchain layer include the DePIN-Specific Chain, tailored specifically to the requirements of DePIN. The IoTeX \cite{IoTex}, leveraging its expertise in IoT domain, provides specialized blockchain solutions for the unique needs of DePIN. On the other hand, the Application Chains facilitates the creation of customized blockchains or subchains within DePIN ecosystem, enabling efficient operations for specific services or data. Additionally, a comprehensive set of blockchain infrastructure tools, such as wallets for managing crypto assets, cross-chain bridges enabling interoperability between different blockchains, and node services providing infrastructural support, form the foundation of blockchain ecosystem. This collection of tools drives the advancement of blockchain technology and promotes its applications. Ultimately, blockchain technology serves as the infrastructural backbone for the governance layer and data layer, for enabling decentralized economics and governance while protecting privacy and security of data in DePINs.

\subsection{Infrastructure layer}
The infrastructure layer forms the cornerstone of the entire architecture, providing the necessary communication, computing, and storage resources for the upper layers. Therefore, the infrastructure layer must possess an extremely high degree of scalability to support the demands of a large-scale DePIN. It consists of the overlay network and the underlay network, covering network devices, computing devices, and terminal devices. Within the overlay network, the peer-to-peer network \cite{p2p} utilizes its decentralized topology and search mechanism to enable efficient communication between nodes without relying on centralized servers. As the underlying network protocol of the blockchain, it provides the decentralized fundamental framework for value transfer and information exchange within DePIN. The underlying network is further subdivided into the core network, the radio access network, and the terminals, which are responsible for the physical transmission of data packets. Devices in terminal devices are particularly critical in DePIN, as they play a central role in data collection, computing, and transmission. Sensors can be deployed in various environments and collect varied information, including temperature, humidity, sound, images, video, and more. The collected data can drive a variety of applications such as navigation maps, energy networks, wireless networks, smart cities, and more. Through wireless communication devices such as WiFi, Bluetooth, etc., DePIN participants can jointly build a decentralized shared network to provide users with diverse network services \cite{wireless}. Data servers provide a large amount of computing and storage resources, which form the hardware foundation of the decentralized data storage system and bandwidth service system in DePINs. The decentralized computation acceleration hardware system provides rentable acceleration hardware (e.g., Graphics Processing Unit, GPU, Tensor Processing Unit, TPU, etc.) for performing complex computational tasks such as machine learning and image processing. 

\section{KEY DESIGN PRINCIPLES}
In this section, we summarize the key principles to consider
when designing the architecture of DePIN.

\textbf{Intelligence}:
To build a more accessible, user-friendly, and efficient DePIN ecosystem, the implementation of intelligent design principles is of paramount importance. This adherence to intelligent design principles not only facilitates a convenient and precise service experience for end-users but also significantly improves the operational efficiency of the entire network. Within the framework of DePIN, interoperability emerges as a pivotal component of intelligent design. It allows seamless communication and interaction between different platforms, applications, and systems, effectively addressing the challenges of data source diversity, data heterogeneity, and consensus protocol compatibility in DePIN ecosystem. Enhanced interoperability \cite{interoperability} enables users to switch between different networks or systems without any hassle, ensuring the consistency and coherence of user identities and assets. It fosters synergistic relationships among diverse services and applications, bringing a more abundant and unified experience to users. This cross-platform and cross-system interoperability is key to realize large-scale DePIN, which drives DePINs toward greater technology integration and a broader user base.

 \textbf{Privacy and security}:
In the context of Web 3.0, security and privacy \cite{Security} are crucial for maintaining the stability of the network and the rights of users, especially in DePIN. The concept of security involves mechanisms for defending against network attacks, hacking, and fraud, to safeguard the robustness of DePIN systems. Privacy, meanwhile, focuses on protecting users' personal information and sensitive data from unauthorized access and misuse. By implementing high standards of security and privacy protection measures, DePIN effectively shields its users from risks such as cyber-attacks, fraudulent practices, and digital theft, thereby building trust and reliability in a decentralized cyber environment. Therefore, to drive widespread adoption of DePIN and active user participation, it is important to prioritize and continuously optimize these protection mechanisms to address evolving cybersecurity threats and privacy protection needs.

 \textbf{Decentralized economy and governance}: 
Decentralized economics and governance together underpin a trustless and democratic network that allows everyone to participate and benefit equally. The decentralized economy presents an innovative economic model that allows users to complete transactions without a trusted authority. Specifically, DePIN adopts a decentralized economic model that incentivizes participants in a permissionless way to share and utilize data or hardware resources, to construct an infrastructure network. Decentralized governance plays a crucial role in ensuring the democracy of the network. It avoids the concentration of power in the hands of a few centralized institutions by enabling all network participants to take part in the decision-making process, thus promoting the active participation and contribution of them. Through this decentralized governance structure, DePIN can achieve broader social participation and more efficient resource allocation.

\textbf{Scalability and low-latency}: 
 Scalability is a measure of a system's ability to process transactions as the network size grows. In DePIN, the network is filled with computational results from physical devices and user transactions, which place extremely high demands on the network's throughput. To support data-intensive applications, and achieve global accessibility, DePIN has to follow a design philosophy that prioritizes scalability to enable seamless user interaction with the networks, while preventing performance bottlenecks and excessive transaction costs. For DePIN, which is built upon numerous physical devices, low-latency communication between devices in the infrastructure layer is also one of the design principles. Emphasizing the connection to the physical world, DePIN applications must ensure the real-time nature of the data sampled from the physical world. Only by ensuring the real-time communication of devices can the network provide a foundation for delivering precise and reliable services to users.

\section{INVESTIGATION OF DePINs}
 In this section, we present a investigation of DePIN, including the applications, the operational mechanism, and the capital market value of DePIN.
\subsection{Applications}
As shown in Fig. \ref{applications}, DePIN applications can be divided into two main categories:
  \begin{figure}
    \centering
    \includegraphics[width=0.5\textwidth]{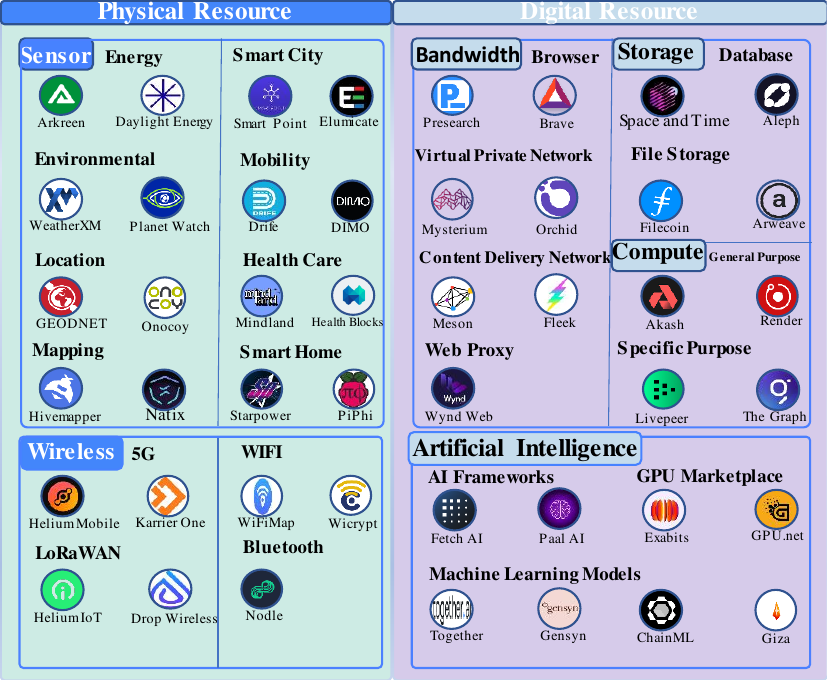}
    \caption{The applications of DePIN can be divided into two main categories: Physical Resource Networks and Digital Resource Networks. Physical Resource Networks and Digital Resource Networks can be further categorized into different application categories.}
    \label{applications}
\end{figure}
\subsubsection{Physical Resource Network} This category of network promotes the deployment of devices that are closely tied to a specific geographic location and that provide unique and irreplaceable services and products based on that location, such as WiFi coverage, environmental monitoring data, and image information. In Physical Resource Networks, participants are rewarded with tokens by contributing valuable data or network resources to the surrounding environment.

Further, Physical Resource Network can be subdivided into two major subcategories, Sensor and Wireless.
\begin{itemize}
\item \textbf{Sensor:} Sensor networks use sensors to collect environmental data in a variety of areas including Energy, Smart City, Environmental, Mobility, Location, Health Care, Mapping, and Smart Home. Energy networks play a key role in monitoring energy consumption and distribution, which is important for promoting sustainability and resource utilization efficiency; Smart City networks collect data on all aspects of city life through the deployment of sensors in urban environments, enabling more efficient city planning; Environmental networks focus on monitoring various aspects of the environment; Mobility networks aim to monitor or provide data on the movement of objects and people, which are critical in applications such as tracking, navigation, and logistics; Location networks utilize Geographical Location technology to measure the object device's precise geographic coordinates; Health Care networks track and monitor personal health status through human wearable sensors, providing valuable data for health management; The Mapping networks focus on capturing geographic and topographic data to create maps and spatial models to support spatial analysis; The Smart Home network enables remote management and intelligent control among smart home devices by connecting physical devices to the blockchain network.

\item \textbf{Wireless:} Wireless networks establish a decentralized network of hotspots through wireless networking hardware devices, including but not limited to 5G, WiFi, Long Range Wide Area Network (LoRaWAN), and Bluetooth. Decentralized 5G network architectures play a crucial role in delivering high-speed, low-latency data transmission services; WiFi networks provide local wireless connectivity and are typically used in public places to provide Internet access within a specific area; LoRaWAN has long range and low power communication characteristics and can provide communication services for IoT networks that require extended battery life and wide range coverage; Bluetooth networks can provide low-range wireless network services to IoT devices and are suitable for short-range communication between wireless devices.
\end{itemize}
 \subsubsection{Digital Resource Networks} This category of network encourages users to share their digital resources, including but not limited to bandwidth, storage space, and computing power. These Digital Resource Networks demonstrate significant competitive advantages over traditional centralized service providers in terms of cost-effectiveness and user accessibility.

Digital Resource Networks are categorized into bandwidth, storage, compute, and AI networks. 
\begin{itemize}
\item \textbf{Bandwidth}: Bandwidth networks can build decentralized media content distribution networks, including Browser, Virtual Private Network (VPN), Content Delivery Network (CDN), and Web Proxy. Browser is the basic tool for users to access and interact with digital resources; Virtual Private Network (VPN) ensures the privacy and security of network communication by providing users with encrypted connections; Content Delivery Network (CDN) effectively reduces the transmission latency of Web services by caching and distributing content across geographically dispersed servers, thereby significantly improving the overall performance of the service; Web Proxy, as an intermediate layer between users and web servers, not only enhances the security and privacy of users surfing the web, but also improves the speed of web.

\item \textbf{Storage}: Storage networks refer to networks that can provide data and file storage, including relational databases and file storage. Relational databases serve as repositories of structured data and provide a solid foundation for data management and manipulation. File storage systems are adept at managing unstructured or semi-structured data, providing an effective method for the storage and retrieval of documents, images, and multimedia files.

\item \textbf{Compute}: Computing networks can provide a variety of computing services, which are categorized into General Purpose and Specific Purpose. General Purpose Computing networks are equipped with the ability to perform diverse computational tasks and provide the necessary computational resources for various applications; Specific Purpose Computing networks, on the other hand, focus on providing specific types of computing resources, such as media files transcoding and data retrieval service.

\item \textbf{AI}: AI networks provide services for the AI industry by aggregating computing resources, including AI frameworks, machine learning models, and GPU marketplaces. Machine learning models can provide pre-trained models for tasks such as natural language processing, image recognition, predictive analytics, etc. AI Frameworks provide a rich set of software libraries and frameworks to support the development and deployment of AI programs; GPU marketplace, a decentralized platform for hardware resource leasing, provides users with leasing services for hardware devices such as GPUs and TPUs from other participants. 

\end{itemize}
\subsection{Operating Mechanism}

\begin{figure*}
    \centering
    \includegraphics[width=0.8\textwidth]{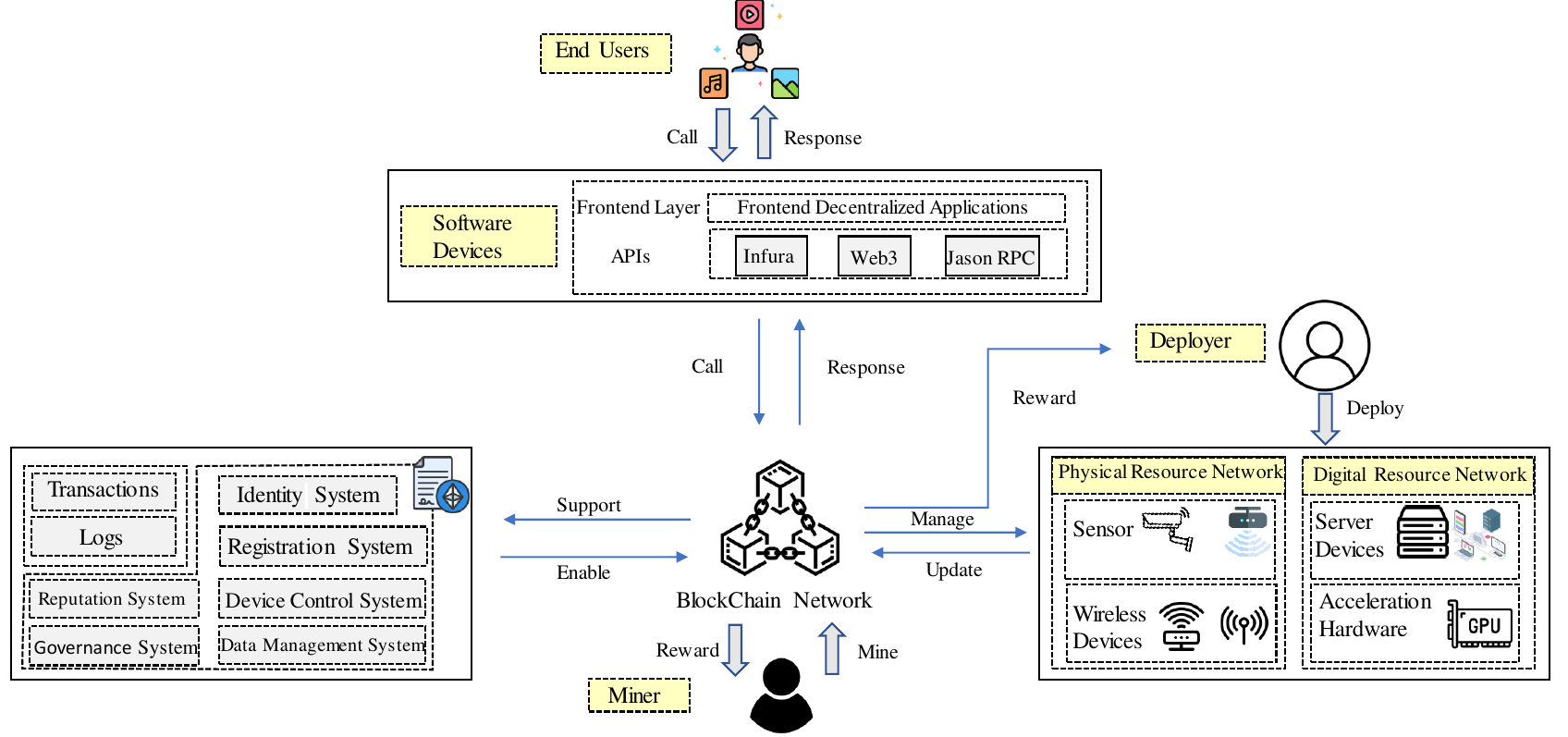}
    \caption{The illustration of the operating mechanism of a  typical DePIN system.}
    \label{workflow}
\end{figure*}

We now briefly introduce the operating mechanism of a typical DePIN system, as illustrated in Fig. \ref{workflow}. 

In DePIN, the physical devices provided by deployers and miners, constitute the underlying infrastructure network. Deployers offer corresponding devices based on the requirements of different infrastructure networks, including sensors, servers, wireless communication devices, and hardware acceleration devices, among others. The devices provided by miners are responsible for validating and packaging blockchain transactions, and ultimately add the blocks that have reached consensus to the end of chain. The resource devices deployed by deployers are managed by the smart contract system on the blockchain network, including the device control system and device data management system. The device control system coordinates hardware resources and allocates hardware devices to users based on user’s transactions. The device data management system manages the data uploaded by deployers' devices and authorizes users access to the corresponding data resources.

Users can purchase services on the frontend of DApps. The purchase transactions are sent to the blockchain networks through the node RPC service, where they await validation and packaging by miners. Confirmed purchase transactions invoke the device control system and device data management system within the smart contract system to obtain the corresponding hardware resources or data services. Ultimately, the fees paid by users are transferred to deployers, who provide the services, as compensation.

\subsection{Market Data}
In this article, we provide a comprehensive analysis of the DePIN capital market for the period of April 2023 to March 2024, covering the three major blockchain platforms Ethereum, solana, and Polygon, on which market capitalization and trading volume data of the various DePIN projects are available, as well as the capital market performance of the DePIN projects.

% \begin{figure}[t]
%     \centering
%     \subfloat[{\small The average monthly market capitalization}]{%
%         \includegraphics[width=0.485\textwidth]{figs/chainsMarketValue.eps}
%         \label{chainsMarketValue}
%     }
%      \hfill % Adds space between the subfigures
%     \subfloat[{\small The trading volume}]{% 使用 \small 命令来缩小字体
%         \includegraphics[width=0.485\textwidth]{figs/chainsTradingVolume.eps}
%         \label{chainsTradingVolume}
%     }
%     \caption{The average monthly market capitalization and the trading volume on Ethereum, Solana, and Polygon.}
%     \label{fig:sidebyside}
% \end{figure}

\begin{figure*}[t]
    \centering
    % 使用 adjustbox 来调整图片的 scale
    \includegraphics[width=\textwidth]{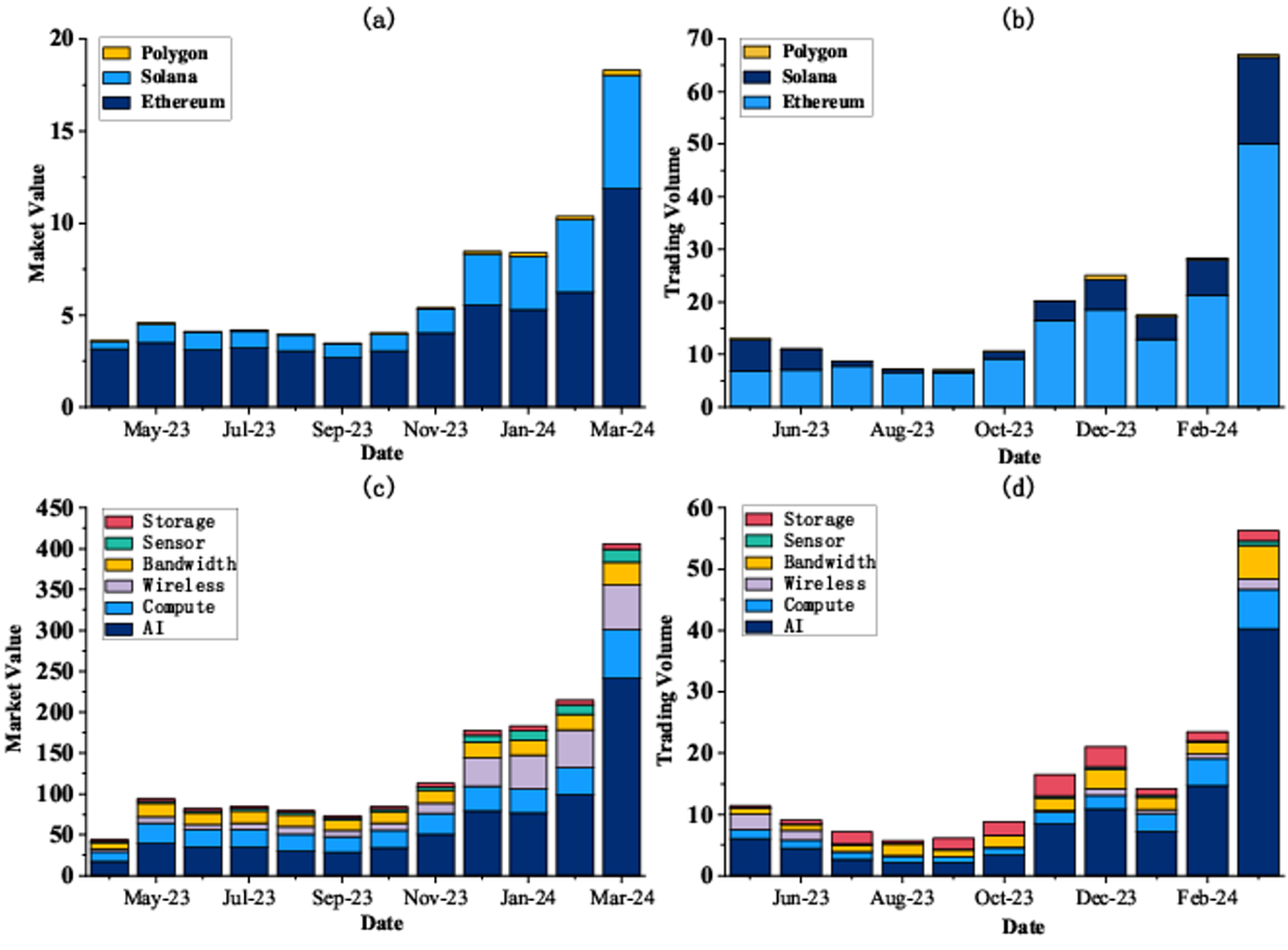}
    \caption{The DePIN projects' market data on Ethereum, Solana, and Polygon: (a) the average monthly market values of the DePIN projects on different blockchains; (b) the average monthly trading volumes of the DePIN projects on different blockchains; (c) the average monthly market values of the DePIN projects under different application categories; (d) the average monthly trading volumes of the DePIN projects under different application categories.}
    \label{marketdata}
\end{figure*}

Fig. \ref{marketdata}(a) shows the average monthly market capitalization of the DePIN projects on Ethereum, Solana, and Polygon. The market capitalization experienced notable growth from April 2023 to March 2024 and surged from  \$3.1\ billion to \$11.8\ billion, reflecting a 326.3\% annual increase. Ethereum hosted the highest DePIN project market capitalization, representing about 64.9\% of the April 2024 average. Solana, however, showed the most significant annual growth at 1,303.6\%, signaling a robust and expanding market. Despite rapid growth, Polygon's market capitalization share was smaller due to a lower starting point. Furthermore, Fig. \ref{marketdata}(b) presents the monthly transaction volumes for the DePIN projects across Ethereum, Solana, and Polygon from May 2023 to March 2024, highlighting a significant annual surge of 646.7\%. This sharp increase underscore the escalating market interest in DePIN. Notably, Solana outperformed with the most impressive growth rate at 707.2\%. Its allure for DePIN initiatives is largely due to its cost-effective transactions, swift processing times, and superior scalability \cite{solana}, positioning it as a favored platform in comparison to Ethereum.

% \begin{figure}[t]
%     \centering
%     \subfloat[{\small The average monthly market capitalization}]{%
%         \includegraphics[width=0.5\textwidth]{figs/categoriesMarketValue.eps}
%         \label{categoriesMarketValue}
%     }
%      \hfill % Adds space between the subfigures
%     \subfloat[{\small The trading volume}]{% 使用 \small 命令来缩小字体
%         \includegraphics[width=0.5\textwidth]{figs/categoriesTradingVolume.eps}
%         \label{categoriesTradingVolume}
%     }
%     \caption{The average monthly market capitalization and the trading volume for six DePIN categories.}
%     \label{fig:sidebyside}
% \end{figure}

Fig. \ref{marketdata}(c) delineates the monthly average market capitalization across the six categories of the DePIN projects on those three platforms from April 2023 to March 2024. The market capitalization exhibited steady trends until October 2023, after the market experienced a notable growth, predominantly within the AI and Wireless categories. By the close of March 2024, AI had claimed a dominant 59.3\% share of the aggregated market capitalization, totaling \$7.7\ billion, underscoring its pivotal influence on market expansion. The projects of Wireless category exhibited substantial growth, establishing both AI and Wireless as the propelling forces behind the dynamic evolution of the DePIN market. In addition, Fig. \ref{marketdata}(d) shows monthly trading volume across the six categories from May 2023 to March 2024. Despite fluctuations, the monthly trading volume had a steady increase. AI projects notably drove trading volume, reflecting AI's strong global investment appeal. Notably, Storage projects, though minor in market capitalization (2.7\%), represented a significant 10.2\% of the trading volume, indicating robust market activity and practical demand.

Synthesizing the data in figures above, it can be concluded that the DePIN market is growing at a rapid rate, with great potential. For academic researchers, these findings can assist them in gaining a deeper understanding of the prominent areas and future development trends within DePIN.

\section{CHALLENGES}
In this section, we discuss the potential challenges that DePIN may encounter across three crucial dimensions: scalability, interoperability, and legality.

\textbf{Scalability}:
 Scalability is a main challenge to blockchain technology \cite{scalability} and is especially closely related to its decentralized nature. DePIN built on blockchain technology is also not immune to such scalability issues. The growth in the number of users and the expansion of the network scale will inevitably increase the volume of transactions in the blockchain network. In particular, the connection between DePIN applications and the physical world imposes higher requirements on information transmission, including the uploading of physical world information and service purchase transactions, which leads to the extension of the transaction confirmation time and the rise of transaction fees.
 
\textbf{Interoperability}:
As shown in the blockchain layer of Fig. \ref{architecture}, DePIN ecosystem is built on top of multiple blockchains. This requires that DePIN applications be able to support homogeneous or heterogeneous state transitions and achieve seamless interoperability with other blockchain networks, which is critical to facilitating efficient communication and data exchange across platforms. while mitigating the problem of cross-chain operations to a certain extent, current interoperability solutions, such as cross-chain bridges based on zero-knowledge proofs or trusted third parties, hooked sidechains, hash locks, etc., are limited to specific blockchain ecosystems or accompanied by high cross-chain costs. 

\textbf{Regulation}:
DePIN, as part of the Web 3.0 ecosystem, also faces multiple regulatory challenges. In terms of regulatory compliance, DePIN's decentralized and anonymous nature makes it more difficult for regulators to monitor the flow of funds. This characteristic may make DePIN ideal for illicit fundraising, pyramid schemes, and money laundering activities. In addition, in terms of tax regulations, DePIN users can be rewarded for their participation in a variety of ways, including by mining, and providing data, hardware, or services. However, due to the anonymous nature of the accounts, it is difficult for the government to collect the evidence required for taxation, which poses a challenge to the existing tax system.

While DePIN technology shows great potential and promise, scalability limitations, interoperability issues, and regulatory barriers remain major obstacles to its widespread adoption. In order to achieve sustainable growth and integration of DePIN technology into mainstream applications, these challenges must be overcome through innovative solutions. 

\section{CONCLUSION}
This article has propose a DePIN architecture, which comprises five distinct layers, namely application layer, governance layer, data layer, blockchain layer, and infrastructure layer. Furthermore, the article delineates four fundamental design principles that underpin the DePIN architecture, including intelligence, privacy and security, decentralized economy and governance, as well as scalability and low-latency. Additionally, the article categorizes the existing decentralized applications within DePIN ecosystem into a three-tiered classification structure. Moreover, the article summarizes a operating mechanism overview of DePIN based on DePIN's applications. Importantly, This article investigates market data of DePIN projects on Ethereum, Solana, and Polygon, offering researchers a clear perspective on the project's growth and potential. Finally, the article introduces the challenges that DePIN faces, including scalability, interoperability, and regulation issues. In conclusion, this article serves as a comprehensive and in-depth survey to DePIN, providing researchers with a solid foundation for comprehending DePIN.

% \begin{thebibliography}{1}
\bibliographystyle{IEEEtran}
\bibliography{ref}
% \bibitem{ref1}
% {\it{Mathematics Into Type}}. American Mathematical Society. [Online]. Available: https://www.ams.org/arc/styleguide/mit-2.pdf

% \bibitem{ref2}
% T. W. Chaundy, P. R. Barrett and C. Batey, {\it{The Printing of Mathematics}}. London, U.K., Oxford Univ. Press, 1954.

% \bibitem{ref3}
% F. Mittelbach and M. Goossens, {\it{The \LaTeX Companion}}, 2nd ed. Boston, MA, USA: Pearson, 2004.

% \bibitem{ref4}
% G. Gr\"atzer, {\it{More Math Into LaTeX}}, New York, NY, USA: Springer, 2007.

% \bibitem{ref5}M. Letourneau and J. W. Sharp, {\it{AMS-StyleGuide-online.pdf,}} American Mathematical Society, Providence, RI, USA, [Online]. Available: http://www.ams.org/arc/styleguide/index.html

% \bibitem{ref6}
% H. Sira-Ramirez, ``On the sliding mode control of nonlinear systems,'' \textit{Syst. Control Lett.}, vol. 19, pp. 303--312, 1992.

% \bibitem{ref7}
% A. Levant, ``Exact differentiation of signals with unbounded higher derivatives,''  in \textit{Proc. 45th IEEE Conf. Decis.
% Control}, San Diego, CA, USA, 2006, pp. 5585--5590. DOI: 10.1109/CDC.2006.377165.

% \bibitem{ref8}
% M. Fliess, C. Join, and H. Sira-Ramirez, ``Non-linear estimation is easy,'' \textit{Int. J. Model., Ident. Control}, vol. 4, no. 1, pp. 12--27, 2008.

% \bibitem{ref9}
% R. Ortega, A. Astolfi, G. Bastin, and H. Rodriguez, ``Stabilization of food-chain systems using a port-controlled Hamiltonian description,'' in \textit{Proc. Amer. Control Conf.}, Chicago, Infrastructure Layer, USA,
% 2000, pp. 2245--2249.

% \end{thebibliography}

\vfill

\end{document}